\RequirePackage{amsmath}
\documentclass[runningheads]{llncs}
\usepackage[T1]{fontenc}
\usepackage{graphicx}

\newcommand{\framework}{\textsc{DAEDalus}}
\newcommand{\papertitle}{\framework: 
Defense Against Firmware ROP Exploits Using Stochastic Software Diversity}

\newcommand{\first}{Selection}
\newcommand{\firstengine}{\first~Engine}
\newcommand{\second}{Diversification}
\newcommand{\secondengine}{\second~Engine}

\newcommand{\semanticequiv}{rewrite}

\newcommand{\ddosim}{DDoSim}

\graphicspath{{figures/}}

\usepackage{tikz}
\newcommand*\circled[1]{\tikz[baseline=(char.base)]{
            \node[shape=circle,draw,inner sep=.7pt] (char) {#1};}}

\usepackage{amsfonts}
\usepackage{comment}

\usepackage{textcomp}
\usepackage{xcolor}

\usepackage{pgfplots}
\usepackage{pgfplotstable}
\pgfplotsset{compat=1.7}

\usepackage{listings}

\usepackage{wrapfig}
\usepackage{float}
\usepackage{atbegshi}
\usepackage{etoolbox}
\usepackage{graphics}
\usepackage{bm}
\usepackage{epsfig}

\usepackage{hyperref}
\usepackage{xurl}

\usepackage[htt]{hyphenat}

\usepackage{cite}

\emergencystretch=\maxdimen

\usepackage{times}
\usepackage{fancyvrb}

\usepackage{booktabs}
\usepackage{dcolumn}
\usepackage[english]{babel}
\usepackage{blindtext}
\usepackage{multirow}

\usepackage{color} 
\usepackage{soul} 
\usepackage{algpseudocode}

\usepackage[linesnumbered,ruled,vlined]{algorithm2e}

\usepackage[bb=ams]{mathalpha}

\usepackage{stackengine}

\usepackage[position=top]{subfig}

\definecolor{mGreen}{rgb}{0,0.6,0}
\definecolor{mGray}{rgb}{0.5,0.5,0.5}
\definecolor{mPurple}{rgb}{0.58,0,0.82}
\definecolor{backgroundColour}{rgb}{0.95,0.95,0.92}

\definecolor{keywords}{RGB}{255,0,90}
\definecolor{comments}{RGB}{0,0,113}
\definecolor{green}{RGB}{0,150,0}
\definecolor{deepgreen}{rgb}{0,0.5,0}

\lstdefinestyle{CStyle}{
    commentstyle=\color{mGreen},
    keywordstyle=\color{blue},
    numberstyle=\tiny\color{mGray},
    stringstyle=\color{mPurple},
    basicstyle=\footnotesize,
    language=C, 
    breakatwhitespace=false,         
    breaklines=true,                 
    captionpos=b,                    
    keepspaces=true,                 
    numbers=left,                    
    numbersep=5pt,                  
    showspaces=false,                
    showstringspaces=false,
    showtabs=false,                  
    tabsize=2
}
\lstdefinestyle{Python3}{
        language=Python, 
        basicstyle=\ttfamily\small, 
        keywordstyle=\color{keywords},
        commentstyle=\color{comments},
        stringstyle=\color{deepgreen},
        showstringspaces=false,
        captionpos=b}

\lstdefinelanguage
   [x64]{Assembler}     
   [x86masm]{Assembler} 
   {morekeywords={CDQE,CQO,CMPSQ,CMPXCHG16B,JRCXZ,LODSQ,MOVSXD, %
                  POPFQ,PUSHFQ,SCASQ,STOSQ,IRETQ,RDTSCP,SWAPGS, %
                  rax,rdx,rcx,rbx,rsi,rdi,rsp,rbp, %
                  r8,r8d,r8w,r8b,r9,r9d,r9w,r9b, %
                  r10,r10d,r10w,r10b,r11,r11d,r11w,r11b, %
                  r12,r12d,r12w,r12b,r13,r13d,r13w,r13b, %
                  r14,r14d,r14w,r14b,r15,r15d,r15w,r15b}} 

\lstdefinestyle{x64asm}{
    commentstyle=\color{mGreen},
    keywordstyle=\color{blue},
    numberstyle=\tiny\color{mGray},
    stringstyle=\color{mPurple},
    basicstyle=\footnotesize,
    language=[x64]Assembler, 
    breakatwhitespace=false,         
    breaklines=true,                 
    captionpos=b,                    
    keepspaces=true,                 
    numbers=left,                    
    numbersep=5pt,                  
    showspaces=false,                
    showstringspaces=false,
    showtabs=false,                  
    tabsize=2,
    morekeywords={CDQE,CQO,CMPSQ,CMPXCHG16B,JRCXZ,LODSQ,MOVSXD, %
                  POPFQ,PUSHFQ,SCASQ,STOSQ,IRETQ,RDTSCP,SWAPGS, %
                  rax,rdx,rcx,rbx,rsi,rdi,rsp,rbp, %
                  r8,r8d,r8w,r8b,r9,r9d,r9w,r9b, %
                  r10,r10d,r10w,r10b,r11,r11d,r11w,r11b, %
                  r12,r12d,r12w,r12b,r13,r13d,r13w,r13b, %
                  r14,r14d,r14w,r14b,r15,r15d,r15w,r15b}
}

\begin{document}
\title{\papertitle}
%
%
\author{Islam Obaidat \and
Meera Sridhar \and
Fatemeh Tavakoli }
\authorrunning{I. Obaidat et al.}
%
\institute{University of North Carolina Charlotte, Charlotte, NC, USA
\email{\{iobaidat,msridhar,ftavakol\}@uncc.edu}}
%
\maketitle              
\begin{abstract}
This paper presents \framework, a software diversity-based framework designed to resist ROP attacks on Linux-based IoT devices. \framework~ generates unique, semantically equivalent but syntactically different rewrites of IoT firmware, disrupting large-scale replication of ROP attacks. \framework~employs \texttt{STOKE}, a stochastic optimizer for x86 binaries, as its core diversity engine but introduces significant extensions to address unique IoT firmware challenges.
\framework's effectiveness is evaluated using \ddosim, a published botnet DDoS attack simulation testbed.
Results demonstrate that \framework~successfully neutralizes ROP payloads by diversifying critical basic blocks in the firmware, preventing attackers from compromising multiple devices for DDoS attacks via memory error vulnerabilities. The findings indicate that \framework~not only mitigates the impact of ROP attacks on individual IoT devices through probabilistic protection but also thwarts large-scale ROP attacks across multiple devices.

\end{abstract}

\section{Introduction}
\label{sec:intro}

Recent works have highlighted the prevalence of \emph{memory error} vulnerabilities in \emph{Internet-of-Things} (IoT) devices~\cite{zhu2019fiot,yu2021towards,chen2021sharing,calatayud2022comparative,serra2022x,nadir2022taxonomy,lee2022ace,zhao2022large}, creating a fertile threat surface for \emph{return-oriented programming} (ROP) attack opportunities on these devices~\cite{luo2020runtime,shi2022harm}. ROP not only enables attackers to perform powerful Turing-complete code manipulations on the compromised devices~\cite{homescu2012microgadgets,clements2017protecting,abbasi2019challenges}, but also allows attackers to create backdoors for compromising large numbers of IoT devices at scale~\cite{obaidat2023ddosim}. Despite these threats, the adoption of certain traditional defense mechanisms, which are often based on \emph{Position-Independent Executable} (PIE) technology, remains low in IoT binaries (below 12\%) due to factors like legacy code and cost implications~\cite{yu2022building,obaidat2023ddosim}. 
Other advanced defenses for ROP attacks on traditional systems, such as \emph{software diversity} (cf.,~\cite{larsen2014sok}) and \emph{control-flow integrity} (CFI) (cf.,~\cite{burow2017control}) are not readily adaptable for IoT systems since they impose high implementation and performance costs, rely on hardware features that are not available in most IoT devices (e.g., segmentation and virtual memory), and require special instructions~\cite{wartell2012binary,nyman2017cfi,almakhdhub2020murai,lee2021savior}. While recent works have explored CFI and software diversity approaches specifically tailored for IoT devices, their associated overheads and deployment constraints prevent them from being used in real-world IoT devices~\cite{wang2020quantitative,tsoupidi2021constraint,spang2021dexie,fu2022fh,christou2022hard,luo2022faslr,shi2022harm}.

We present \framework~(\underline{D}efense \underline{A}gainst firmware ROP \underline{E}xploits using stochastic software \underline{D}iversity), a software diversity-based framework for thwarting ROP attacks on x86 Linux-based IoT devices. \framework~creates unique semantically equivalent, but syntactically different, copies of input firmware binaries by changing their internal structure~\cite{zhang2021diversity}. The main insight behind our approach is that ROP attacks require precise knowledge about the binary implementation of the target firmware, such as the exact memory address of code snippets that remove items from the stack and push them to specific registers, to construct the attack payload. Having semantically equivalent, but syntactically different binaries, which we call \emph{\semanticequiv(s)}, makes it challenging to conduct an ROP attack because an attacker has no knowledge about the binary implementation of each \semanticequiv. Furthermore, \framework~discourages large-scale ROP attacks on IoT devices since the attacker has to perform painstaking attack payload customization for each \semanticequiv. 

We employ \texttt{STOKE}~\cite{schkufza2013stochastic} as \framework's core diversity engine. \texttt{STOKE} is a stochastic optimizer and synthesizer for x86 binaries; while \texttt{STOKE} is designed to find the single, \emph{most performant} \semanticequiv, \framework's goal is \emph{security}---to find multiple \semanticequiv{s} of the original binary to detract the attacker. This fundamental difference in goal requires major modifications and enhancements to \texttt{STOKE}'s design and implementation. \texttt{STOKE}'s prototype~\cite{stoke_git} uses function-level \semanticequiv~processing and production for a given binary. In \framework, we adapt \texttt{STOKE} to process \emph{basic block}-level \semanticequiv{s}~for an input IoT firmware binary instead of functions. Particularly, IoT devices, constrained by limited resources and specific functional requirements, often operate using small, optimized instruction sets. Thus, granular control over \semanticequiv{s} is crucial to ensure the firmware's functionality remains intact. Basic blocks provide this granularity, ensuring deterministic behavior of the \semanticequiv~instructions and minimizing the risk of inadvertently introducing malfunctions. 


\texttt{STOKE} requires manual provision of a list of registers and memory locations that the input function accesses and modifies. In \framework, we automate this by adding a localized data flow analysis that uses \emph{data dependencies} between variable definitions (registers and memory locations) of a given basic block. Also, \texttt{STOKE} uses random test cases to validate generated \semanticequiv{s}, which may be suboptimal, particularly for identifying hidden branches or edge cases in the code. To address this limitation, we develop an integration with \texttt{AFL++} fuzzer~\cite{fioraldi2020aflpp} for targeted test case generation. This enhances the robustness of the \semanticequiv~generation process in \framework~ since our test cases are generated using \texttt{AFL++}'s genetic algorithms, which cover a broader range of the program's state space compared to simple random test cases.

In general, a successful ROP attack on target software stems from the availability of basic blocks that contain: (a) code snippets from the target software that the attacker can leverage to construct the ROP payload and (b) a memory error vulnerability that acts as an entry point (i.e., backdoor) to this target software~\cite{homescu2012microgadgets,goktacs2014size,english2019exploiting}. We define basic blocks that contain one of (a) or (b) above as \emph{security-critical} basic blocks. Past work has demonstrated that diversifying \emph{only} security-critical basic blocks suffices for mitigating ROP attacks~\cite{pappas2012smashing}. Furthermore, diversifying all basic blocks in the target program can be impractical due to high performance penalties and other reasons (e.g., network packet protocols are often unsuited for instrumentation due to protocol conflicts~\cite{kim2019polar}). Towards this, we systematically define and categorize what exactly constitutes security-critical basic block diversification candidates and develop algorithms to automatically identify them in a target binary, and only diversify these basic blocks in that target.

For evaluating \framework, we utilize \emph{\ddosim}, a published botnet \emph{distributed denial-of-service} (DDoS) attack simulation framework~\cite{obaidat2023ddosim}. \ddosim~constructs attack scenarios using Docker containers to create simulated IoT devices loaded with actual IoT binaries, uses ROP payloads to exploit memory error vulnerabilities in these devices remotely to gain backdoor access, and creates DDoS attacks on a simulated target server with the compromised devices over an NS-3 simulated network. In our work, we construct an ROP payload for each input binary, diversify a subset of security-critical blocks in each input binary, deploy our input target binary and its synthesized rewrites on \ddosim~and replicate the attack scenario used in \ddosim. Our chosen basic block subset specifically comprises all basic blocks containing ROP gadgets that an attacker uses to manipulate register values by sourcing these values from the attack payload (e.g., from the stack in stack-based buffer overflows).
Our evaluation results indicate that \framework~effectively neutralizes the ROP payload, and prevents an attacker from using a single attack payload to mount a DDoS attack via memory error vulnerabilities.

\begin{figure*}[ht]
\centering
\includegraphics[width=0.98\linewidth]{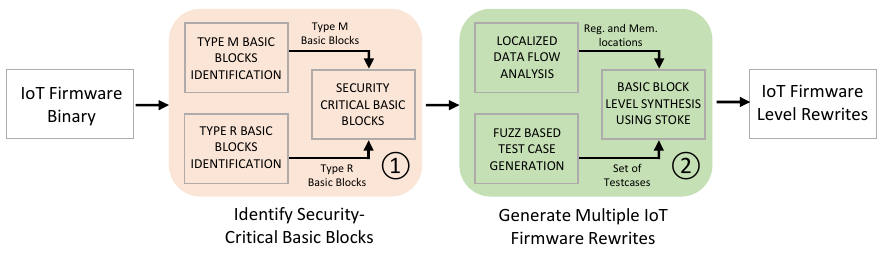}
\caption{\framework~ Workflow}
\label{fig:mainframework}
\end{figure*}

While \framework~is also applicable to traditional (non-IoT) systems, it is specifically designed for IoT firmware. Firstly, \framework~operates at the basic block granularity level, allowing for tractable and verifiable changes. This granularity is crucial as alternate levels, such as function-level granularity, can be challenging to delineate in binaries and may inadvertently introduce malfunctions in IoT firmware~\cite{andriesse2016depth,wang2017ramblr,salehi2020musbs}. \framework~selectively diversifies only security-critical basic blocks, and not the entire binary, and creates rewrites that are equal to or smaller in size than the original blocks; these features align well with the resource constraints of IoT devices. \framework~also operates at the binary level, as opposed to other solutions that often necessitate source code access, making these solutions inapplicable to the majority of IoT firmware (which is often proprietary)~\cite{tsoupidi2023thwarting}. \framework's offline application addresses operational constraints unique to IoT devices, unlike other solutions that require additional libraries or VMs to randomize the binary for each run~\cite{andriesse2017compiler,andriesse2016depth}. Defenses such as PIE depend on runtime features such as ASLR, which may not be compatible with IoT systems~\cite{yu2022building,obaidat2023ddosim}.

In addition, \framework's diversity approach is a valuable security measure for IoT firmware specifically since IoT firmware are typically deployed at massive scale, many times in consumer or safety-critical devices~\cite{obaidat2023ddosim}. The same firmware, with the same vulnerability, replicated at scale, makes them a perfect target for large-scale DDoS attacks. To our knowledge, no other solution is specifically tailored to address large-scale memory error exploits in IoT devices with the same efficiency as \framework.

To summarize, the main contributions of our work include:
\begin{itemize}
    \item We develop \framework, a framework that uses software diversity to thwart ROP attacks targeting IoT firmware binaries without imposing significant overhead or requiring special hardware features and extensions, which makes it suitable for the IoT domain. While \framework~is built upon \texttt{STOKE}~\cite{schkufza2013stochastic}, a stochastic superoptimizer, we make significant modifications to, and add extensions to, \texttt{STOKE}, tailored for the goal of security rather than optimization. 
    \item We develop a technique for converting the default function-level processing in \texttt{STOKE} to basic block-level processing. Our algorithm handles conversion of various termination instructions of basic blocks such as calls and jumps into \texttt{ret} instructions, which can be error-prone for producing semantically-equivalent rewrites that match all control flows of the original binary. 
    \item We add a localized data flow analysis for an individual basic block, to automate the process of input generation (registers and memory locations accessed and modified by the basic block) for rewrite synthesis. 
    
    \item We develop a technique for integrating AFL++ fuzzing with \texttt{STOKE} for \emph{focused} test case generation, for verifying synthesized rewrites, to replace the default random test case generation. 
    
    \item We define and categorize security-critical basic blocks from a given firmware binary and develop algorithms to automatically identify them. \framework~only diversifies these basic blocks instead of diversifying the entire IoT firmware binary.
    \item We evaluate \framework~using \ddosim, an open-source realistic simulation environment. In this setting, our experimental results demonstrate that at least two basic block-level \semanticequiv{s} can effectively prevent attackers from leveraging ROP payloads to gain unauthorized access to target IoT devices.
\end{itemize}

\paragraph{\framework~ Overview.} Fig.~\ref{fig:mainframework} presents \framework's workflow.
In the security-critical basic-block selection phase (\circled{1}), \framework~ takes as input an IoT firmware binary and produces sets of security-critical basic blocks as output. The next phase (\circled{2}), \framework~ takes the produced set as input and generates basic block-level \semanticequiv{s} for each basic block in this set. Then, \framework~ replaces the security-critical basic blocks with basic block-level \semanticequiv{s} in the input binary and generates different firmware \semanticequiv{s}, where each firmware \semanticequiv~has one or more basic block-level \semanticequiv{s}.
Once we generate \semanticequiv{s}~for an input firmware binary, we deploy these binary \semanticequiv{s} on the \ddosim~ simulation framework~\cite{obaidat2023ddosim} to verify the effectiveness of \framework~in thwarting ROP attacks. 
 Section~\ref{sec:selection} describes how we define and identify security-critical basic blocks. Section~\ref{sec:synthesis} discusses our diversification process. Section~\ref{sec:evaluation} details the evaluation of \framework. Section~\ref{sec:related} presents related work, and~\S\ref{sec:conclusion} presents conclusions.

\paragraph{Adversarial Model.} In our work, we assume that the memory regions in the target IoT devices are only writable or executable, but not both (i.e., W$\oplus$X is enabled). Hence, the attacker does not have the ability to conduct \emph{code injection} attacks (in these attacks, the attacker injects the malicious code into the program's memory and redirects the program's execution flow to this injected code) and can perform ROP attacks. Furthermore, we consider that the attacker has access to the \emph{original equipment manufacturer}'s (OEM) firmware binary to perform any analysis required to conduct an ROP attack (the attacker can purchase an IoT device and extract the binary from this device or can obtain it online if available). Also, we do not consider the category of \emph{non-control data} attacks~\cite{szekeres2013sok} (e.g., \emph{Data-Oriented Programming}~\cite{hu2016data}). In these attacks, the attacker modifies critical data in the program's memory, which controls the behavior of this program. Consequently, the execution of the program changes according to these modifications.

\section{Security-Critical Basic Block Selection}
\label{sec:selection}

Manual identification of the security-critical basic blocks is infeasible in practice due to the large number of basic blocks in a given firmware binary. For example, \texttt{Connman}, a network connection manager used widely in IoT devices~\cite{english2019exploiting}, has 41857 basic blocks (obtained using \texttt{angr}~\cite{shoshitaishvili2016state}, a framework for analyzing binaries). Therefore, in \framework, we focus on two important goals in the \firstengine: (i) defining and categorizing what exactly constitutes a security-critical basic block, and (ii) developing an automated technique for identifying them. 


\paragraph{Type R Basic Blocks.}
Type R basic blocks may contain code snippets from the target binary that an attacker uses to construct the ROP payload. ROP attacks subvert the legitimate control-flow execution of the compromised system to its malicious payload. A typical ROP payload utilizes multiple \emph{gadgets}~\cite{shacham2007geometry,dhavlle2022cr}, fragments of code already available in the victim's memory to perform simple tasks (e.g., copy a string to a specific memory address), chained together to achieve the attacker's goal (e.g., spawn a shell process). Once an attacker redirects the legitimate control-flow execution to the first gadget in the ROP  payload, this gadget performs its required task and then transfers the execution flow to the next gadget in this chain. Similarly, all gadgets in this chain get executed in the attacker's intended order for the attack to succeed. Any change to these gadgets renders the attack ineffective.


Diversifying the basic block that contains an ROP gadget changes the instructions in this block, which may result in different instructions that do not perform the gadget's task~\cite{pappas2012smashing}. \framework~uses software diversity in an attempt to modify (and eliminate) as many ROP gadgets available in the firmware binary as possible. This process breaks the gadgets' semantics in the IoT firmware binary without affecting the semantics of the actual firmware binary, rendering the ROP payload ineffective.

We first categorize ROP gadgets to be able to identify all gadgets that are more likely to be used in ROP attacks against an IoT firmware binary. Prior works classify ROP gadgets into three main classes: \emph{Change Memory}, \emph{Change Register}, and \emph{Call} gadgets~\cite{schwartz2011q,shoshitaishvili2016state}. We follow this classification to identify gadgets in a given IoT firmware binary. The first class, \texttt{Change Memory}, writes a value to a memory address to manipulate a set of instructions or a function in the target IoT binary. For example, an attacker can write a \texttt{url} string to the target's BSS memory using \texttt{Change Memory} gadget to later download a malware file from this \texttt{url}. The \texttt{Change Register} class changes the value of a given register in a target IoT firmware binary. For example, an attacker writes a value to a specific register using a \texttt{Change Register} gadget to provide it as an argument to a function (x86\_64 calling convention uses registers to pass arguments to functions~\cite{lu2018system}). The \texttt{Call} class consists of gadgets that perform a call to a function (e.g., calling libc \texttt{system()} function).
In \framework's implementation, we utilize \texttt{angr}, \texttt{Radare}\url{https://rada.re/n/radare2.html} (a tool that assists in disassembling, debugging, analyzing, and manipulating binaries), and \texttt{Ropper}\url{https://github.com/sashs/Ropper} (a tool that finds ROP gadgets in binaries) tools to identify Type R basic blocks in the firmware. 


To create a Type R basic block list in \framework, we begin by constructing a \emph{Control-Flow Graph} (CFG) of the input firmware binary using \texttt{angr}. We then use this CFG to scan the firmware for all \texttt{Call} gadgets (which are represented as \texttt{call} instructions that can be located on some of the edges in the constructed CFG) that can be leveraged in ROP attacks (e.g., \texttt{call} \texttt{execlp()} can be utilized in ROP to start a process). We also scan the input binary using both \texttt{Radare} and \texttt{Ropper} to identify \texttt{Change Register} and \texttt{Change Memory} gadgets.

Once we collect all required ROP gadgets, we use the addresses of these gadgets to locate Type R basic blocks. Using the CFG nodes, we check if the address of an ROP gadget corresponds to a basic block's address or if it falls within a basic block boundary. In either case, the identified basic block is added to the Type R list.

\paragraph{Type M Basic Blocks.}
Type M basic blocks are basic blocks that both (i) contain function calls to external libraries that may contain vulnerabilities attackers can leverage to mount ROP attacks, such as \texttt{memcpy} and \texttt{strcpy} \texttt{libc} functions, and (ii) use these library functions to process input data originating from external sources into the IoT firmware binary. This specification is essential to note, as not every call to an external library function inherently poses a risk, even when handling unsanitized data. 
Diversifying Type M basic blocks is important because diversification can alter execution flow and stack structure while maintaining the same semantics as the original code. Yet, the new execution flow and stack structure can disrupt the contiguity of memory locations that an ROP exploit may otherwise leverage, thus countering the ROP payloads designed to exploit the original assembly sequence. 

For implementation, we begin by collecting the addresses of all calls to library functions by searching for \texttt{call} instruction in the CFG of the binary (generated using \texttt{angr}). Then, we perform a taint analysis to identify all code locations that affect the inputs of these library function calls. We implement our static taint analysis engine using \texttt{angr}, as suggested in~\cite{gritti2021reaching,soong2021angrtaint}. The suggested approach leverages \emph{Reaching Definition Analysis} (RDA), a static analysis approach in \texttt{angr} to identify the definitions (registers, memory or code locations) that affect a variable at a particular observation point in a specific function~\cite{maso2020reachingdef}, to track all changes done to a particular variable.

Since \texttt{angr} performs RDA on a single function (not over multiple functions), we start the RDA from the \texttt{main} function, and then we recursively create child RDAs for each \texttt{call} instruction until we reach our observation point. Thus, in our work, we set the observation point at the address of a library function call (address of \texttt{call} instruction), and we run the RDA on the \texttt{main} function of the IoT firmware. Then, every time we encounter a \texttt{call} instruction to a function (inside the \texttt{main} function), we start a new \emph{child} RDA on this new function. Once \texttt{angr} finishes the analysis on the new function, it copies the analysis state back to the \texttt{call} instruction that forked the child RDA to continue executing the parent RDA until it reaches the required observation point.

Based on the RDA results, we determine whether a library function processes data from outside the firmware (e.g., arguments provided to the \texttt{main} function). If the library function processes data from untrusted sources, we add the address of the basic block that contains the \texttt{call} instruction of this library function to the Type M list.

\section{Rewrites Synthesis}
\label{sec:synthesis}


Our \secondengine~(see Figure~\ref{fig:mainframework}) aims to produce \texttt{N} basic block-level \semanticequiv{s} (\texttt{N} is a user-specified number) for Type R and Type M basic blocks (obtained from our \firstengine). This engine's core component is \texttt{STOKE}, which requires three main inputs: (i) a file that contains the assembly instructions of the target function, (ii) a set of registers and memory locations accessed and modified by this function, and (iii) a set of test cases. To enable automatic \semanticequiv{s} production at the basic block level in the \secondengine, we perform several modifications and enhancements to each of \texttt{STOKE}'s three inputs, which are all discussed in detail below. 



\paragraph{Basic Block-level Synthesis.}

As explained in~\S\ref{sec:intro}, for several practical reasons we choose to diversify the input binary at the basic block level, as opposed to \texttt{STOKE}'s default process. 
Since \texttt{STOKE} takes a function as input (and produces multiple function-level \semanticequiv{s}), we represent a basic block as a function and then pass it to \texttt{STOKE}. 
Functions end with a \texttt{ret} instruction, but basic blocks typically terminate with \emph{calls}, \emph{unconditional jumps}, or \emph{conditional jumps}. We change these end instructions of basic blocks into \texttt{ret} instructions. 
For cases in which basic blocks end with calls or unconditional jumps, we also change these instructions to \texttt{ret} instructions.

\newsavebox{\basicblockimAsm}
\begin{lrbox}{\basicblockimAsm}
\begin{lstlisting}[style=x64asm,escapeinside={(*}{*)}]
add 4,%eax
test %eax,%eax
je 0x112346608
\end{lstlisting}
\end{lrbox}

\newsavebox{\modblockimAsm}
\begin{lrbox}{\modblockimAsm}
\begin{lstlisting}[style=x64asm,escapeinside={(*}{*)}]
add 4,%eax
test %eax,%eax
ret
\end{lstlisting}
\end{lrbox}

\newsavebox{\rewriteAsm}
 \begin{lrbox}{\rewriteAsm}
\begin{lstlisting}[style=x64asm,breaklines=true,escapeinside={(*}{*)}]
lea 4(%eax),%eax
ret
\end{lstlisting}
 \end{lrbox}

\begin{figure}[ht]
\centering
  \subfloat[]{\label{lst:basic_block_asm} \usebox{\basicblockimAsm}} \hspace{0.4in}
  \subfloat[]{\label{lst:basic_block_change_asm} \usebox{\modblockimAsm}} \hspace{0.4in}
  \subfloat[]{\label{lst:rewrite_asem} \usebox{\rewriteAsm}}
  \caption{Example of diversification error in conditional jump} \label{fig:simple_block_je_example}
\end{figure}

In the conditional jumps, changing these exit instructions to a \texttt{ret} instruction in input basic blocks may result in an error. Here, a produced basic block-level \semanticequiv~may maintain the correct functionality for only one execution path of the input basic block and ignore the other path. For e.g., Listing~\ref{lst:basic_block_asm} shows a basic block with three assembly instructions, \texttt{add}, \texttt{test}, and \texttt{je}. The first instruction increments \texttt{eax} by 4. The second instruction, \texttt{test}, performs a bit-wise logical \texttt{AND} between the first operand (\texttt{\%eax} register) and the second operand (\texttt{\%eax}) without affecting the values stored in these operands. Based on the result of the \texttt{AND}, the \texttt{test} instruction sets the \texttt{SF}, \texttt{ZF}, and \texttt{PF} status flags in the \texttt{RFLAGS} register. The conditional jump (\texttt{je}) branches to its destination (i.e., 0x112346608) if the value of the \texttt{ZF} flag is equal to 1, and otherwise, it continues the sequential execution of the instructions that come after it.

In this example, changing \texttt{je} to \texttt{ret} results in the basic block shown in Listing~\ref{lst:basic_block_change_asm}. Providing this modified basic block to \texttt{STOKE} may produce the \semanticequiv~shown in Listing~\ref{lst:rewrite_asem}, which is functionally equivalent to the basic block in Listing~\ref{lst:basic_block_change_asm}. This \semanticequiv, however, is only functionally equivalent to one execution path in the original basic block, which always executes the instructions that come after this basic block (in a sequential manner) without ever branching to 0x112346608.

This issue occurs since, unlike the input basic block, the instructions that precede the conditional jumps in the produced basic block-level \semanticequiv{s} may not affect the \texttt{RFLAGS} register (which controls the destination of the jump instruction) in the same manner. We can resolve this issue by forcing \texttt{STOKE} to always check the value of the \texttt{RFLAGS} register when producing \semanticequiv{s}, which is not checked by default in \texttt{STOKE}. However, always checking this value 
limits the number of possible \semanticequiv{s} for an input basic block. For instance, suppose we process a basic block that has an \texttt{add} instruction followed by an unconditional \texttt{jmp} instruction. This \texttt{add} instruction impacts the value of the \texttt{RFLAGS} register (most ALU instructions, such as \texttt{sub} and \texttt{add} instructions, change this value~\cite{guide2011intel}). The impact of the \texttt{RFLAGS} register may not be important in this scenario since it is followed by an unconditional jump instruction (\texttt{jmp}). In this example, checking the value of the \texttt{RFLAGS} register when producing \semanticequiv{s} limits the number of possible \semanticequiv{s} since the instructions of a \semanticequiv~candidate have to perform the required functionality (i.e., performing the addition) and change the value of the \texttt{RFLAGS} register in the same manner as the input basic block. To overcome this limitation, we only instruct \texttt{STOKE} to check the value of the \texttt{RFLAGS} register when processing an input basic block that ends with a conditional jump instruction.

In order to perform the above changes, we create an automated script that does the following. First, it changes the exit instruction of input basic blocks to a \texttt{ret} instruction. Second, it configures \texttt{STOKE} to check the \texttt{RFLAGS} register before passing an input basic block with an unconditional exit instruction. Third, this script represents each input basic block in \texttt{STOKE}'s format~\cite{stoke_git} since \texttt{STOKE} expects the input to follow a specific format (containing the assembly instructions, directives, and labels). Fourth, in order to maintain the original binary size, we instruct \texttt{STOKE} to generate a \semanticequiv~that is equal to or smaller than the size of the original basic block. This is crucial in preventing any potential overhead associated with the size of the modified firmware binaries. Fifth, this script invokes \texttt{STOKE} with our modified input basic block and the customized configuration, which allows \texttt{STOKE} to produce one \semanticequiv~for this basic block. Sixth, this script can loop to generate N basic block-level \semanticequiv{s} as per the user's preference. Lastly, this script reverses the \texttt{ret} instruction of each produced \semanticequiv~back to the original exit instruction of the input basic block.


\paragraph{Localized Data Flow Analysis.}
\texttt{STOKE} lacks a built-in capability for automatically providing a list of registers and memory locations accessed and modified by an input basic block, a prerequisite for \texttt{STOKE}'s \semanticequiv~synthesis. To address this gap, we implement an automated localized data flow analysis feature into \framework. This specialized approach focuses on data flow within specific regions of a program~\cite{wang2017ramblr}. The need for such localized analysis becomes especially pertinent when considering the intricate nature and scale of IoT firmware binaries. For instance, the localized nature of our analysis circumvents the computational limitations of performing data flow analysis on an entire IoT firmware binary, which requires considerable computational resources that increase with the size of the firmware binary. Furthermore, our automated methodology cuts down on the manual labor previously necessary for input required by \texttt{STOKE}, which not only speeds up the \semanticequiv~synthesis process but also minimizes the likelihood of human error. By localizing the data flow analysis, we are able to mitigate these complexities and ensure that the analysis remains tractable for IoT firmware binaries.

We perform \emph{static} localized data flow analysis on an input basic block to determine data dependencies among registers and memory locations. 

In \framework, to incorporate static localized data flow analysis, we start by delineating the boundaries of the basic block using the CFG generated by \texttt{angr}. 
Then, we employ \texttt{angr}'s data dependency graph constructor to establish the graph corresponding to this basic block. Here, the graph construction is achieved in tandem with the generated CFG. After assembling the data dependency graph for the basic block, we use it to extract the required list of registers and memory locations that are accessed or modified. This list, along with the target basic block, is subsequently fed into \texttt{STOKE}, automating its synthesis process and enhancing both its speed and accuracy.


\paragraph{Fuzz-Based Test Case Generation.}
\texttt{STOKE} relies on test cases for detecting incorrect function-level \semanticequiv{s}. \texttt{STOKE} uses PinTool~\cite{luk2005pin}, a dynamic binary instrumentation framework for x86, to create a custom execution environment (EE) and populates the registers and memory locations accessed by an input function using these test cases. \texttt{STOKE} executes a compiled version of the input function in this EE, generating an output represented as modifications to the EE's values. \texttt{STOKE} then compares this generated output with similar output values generated from executing a \semanticequiv~candidate in the same EE and, based on the comparison, assesses the \semanticequiv~candidate's correctness.

\texttt{STOKE} offers several methods for generating test cases for an input function, with random generation being the default method. However, random generation may not always produce optimal test cases, as we demonstrate using our example in Listing~\ref{lst:fuzz_motivate}. This listing shows assembly instructions consisting of the \texttt{sub}, \texttt{add}, and \texttt{test} instructions, which respectively decrement the \texttt{\%rax} register, increment the \texttt{\%rdx} register, and perform a bitwise logical AND operation on the \texttt{\%rdx} register to set the \texttt{SF}, \texttt{ZF}, and \texttt{PF} status flags. In this example, the \texttt{test} instruction sets \texttt{OF} and \texttt{CF} flags to 0 and the \texttt{ZF} flag to 1 when the value of the \texttt{\%rdx} register is zero. For this simple example, we discuss two possible scenarios in which \texttt{STOKE} may generate unreliable test cases:

\newsavebox{\motAsOddReg}
\begin{lrbox}{\motAsOddReg}
\begin{lstlisting}[style=x64asm]
 sub $0x1, %rax
 add $0x1, %rdx
 test %rdx, %rdx
\end{lstlisting}
\end{lrbox}

\newsavebox{\motAsOddSyn}
\begin{lrbox}{\motAsOddSyn}
\begin{lstlisting}[style=x64asm]
 inc %rdx
 dec %rax
\end{lstlisting}
\end{lrbox}

\begin{figure}[ht]
\centering
  \subfloat[]{\label{lst:fuzz_motivate} \usebox{\motAsOddReg}} \hspace{0.3in}
  \subfloat[]{\label{lst:fuzz_motivate_rewrite} \usebox{\motAsOddSyn}}
  \caption{Example of an unreliable test case generated randomly} \label{fig:fuzz_example}
\end{figure}

\textbf{Scenario 1:} \texttt{STOKE}'s random test case generation may produce test cases that set the values of the \texttt{OF} and \texttt{CF} to 0 in the custom execution environment. When executing the \texttt{test} instruction with these cases, both \texttt{OF} and \texttt{CF} remain unchanged at 0, suggesting the \texttt{test} instruction has no effect on these flags. As a result, \texttt{STOKE} may disregard the \texttt{test} instruction during synthesis, as it appears to have no impact on input/output test cases. Consequently, \texttt{STOKE} generates the \semanticequiv~shown in Listing~\ref{lst:fuzz_motivate_rewrite} that is only functionally equivalent to the first two instructions of the original code in Listing~\ref{lst:fuzz_motivate}.
    
\textbf{Scenario 2:} Typically, the \texttt{test} instruction is followed by a conditional jump instruction (such as \texttt{je}, which jumps to its target if the \texttt{ZF} flag is 1). For the \texttt{test} instruction to set the \texttt{ZF} flag to 1, the \texttt{rdx} register value must be zero. However, \texttt{STOKE}'s random test case generation has a low probability of setting the \texttt{rdx} register to zero (nearly 1/$2^{64}$). Consequently, most randomly generated test cases have non-zero \texttt{rdx} values, resulting in a \texttt{ZF} flag value of 0 after executing the \texttt{test} instruction in the custom environment. Again, if the generated test cases yield a \texttt{ZF} flag value of 0 before and after executing the \texttt{test} instruction, \texttt{STOKE} perceives the instruction as having no effect on the \texttt{ZF} flag and may ignore it during the synthesis process.

Our example above demonstrates how \texttt{STOKE}'s random test case generation could impact the identification of incorrect \semanticequiv~candidates during the synthesis process. Other mechanisms provided by \texttt{STOKE} for generating test cases require the user to manually write custom test cases, use symbolic execution that requires the input function to satisfy certain criteria, or use a buggy dynamic instrumentation that does not work properly~\cite{stoke_git}. Due to limitations in these other mechanisms, we employ the \texttt{AFL++} fuzzer~\cite{fioraldi2020aflpp} to generate targeted test cases for basic blocks. \texttt{AFL++} fuzzer utilizes \texttt{QEMU}~\cite{bellard2005qemu} to instrument binaries and record branch information during execution. The fuzzer runs the target program with malformed or semi-malformed inputs, records branch coverage information, and uses inputs that lead to new branches in a genetic algorithm for generating subsequent inputs. This process enables \texttt{AFL++} to observe transitions between basic blocks and to collect data on which branches in the target program are exercised. In the remainder of this subsection, we elaborate on the use of \texttt{AFL++} for generating test cases for input basic blocks.

To generate test cases for a specific basic block using \texttt{AFL++}, \framework~integrates the assembly instructions of this basic block into pre-prepared assembly code, which we refer to as \emph{fuzz assembly}. The resulting assembly is then compiled into a binary that we fuzz using \texttt{AFL++}. Our fuzz assembly performs two main tasks:

\textbf{Input Redirection}: The fuzz assembly redirects the fuzzing input (generated by \texttt{AFL++}) to the registers that typically hold the default arguments of a function. Specifically, it redirects the fuzzing input to either pointer/integer argument registers (\texttt{\%rdi}, \texttt{\%rsi}, \texttt{\%rdx}, \texttt{\%rcx}, \texttt{\%r8}, and \texttt{\%r9}) or floating-point argument registers (\texttt{\%xmm0}, \texttt{\%xmm1}, \texttt{\%xmm2}, \texttt{\%xmm3}, \texttt{\%xmm4}, \texttt{\%xmm5}, \texttt{\%xmm6}, and \texttt{\%xmm7}). If a basic block expects input in a specific register (e.g., \texttt{\%rax}), we insert a \texttt{mov} instruction at the beginning of this basic block (e.g., \texttt{mov} \texttt{\%rdi,\%rax}), effectively redirecting the fuzz input from \texttt{\%rdi} to the required register. If the basic block expects a value in a specific memory address, we can use other similar instructions to redirect the input (e.g., \texttt{mov} \texttt{\%rdi,(\%rax)}, assuming that this basic block expects an input value stored in a memory address in \texttt{rax}). Moreover, if the block expects input values in multiple registers or specific memory locations, we can redirect them using multiple appropriate \texttt{mov} instructions. We identify the set of registers and memory locations that are used as input for a basic block using information obtained from our data dependency analysis).

\textbf{Output Redirection}:  At the end of each basic block, the fuzz assembly redirects the flow of execution to the \texttt{abort@PLT} \texttt{libc} function. This action ensures that \texttt{AFL++} records unique crashes resulting from the fuzz inputs since \texttt{AFL++} is good at mutating its input and storing the inputs that cause unique crashes. If the basic block ends with a \texttt{ret} instruction, we replace it with a \texttt{call} \texttt{abort@PLT} instruction, thereby ensuring that the \texttt{abort} function triggers a crash that \texttt{AFL++} can detect. Similarly, we replace all unconditional jumps and calls within basic blocks with \texttt{call} \texttt{abort@PLT} instructions. On the other hand, for basic blocks ending in conditional control-flow transfer instructions (e.g., \texttt{je}), we modify the destination to the abort function (e.g., change \texttt{je} \texttt{0x401080} to \texttt{je} \texttt{abort@PLT}) and append a \texttt{call} \texttt{abort@PLT} instruction immediately after the conditional jump. This approach enables \texttt{AFL++} to detect both execution paths of the conditional jump as unique crashes.

To generate test cases using \texttt{AFL++}, we create an automated script that performs the following steps. The script includes the assembly instructions of an input basic block into our fuzz assembly. It uses the compiler available with \texttt{AFL++} to generate a binary from the modified fuzz assembly. The script invokes \texttt{AFL++} to fuzz the generated binary, which results in recording the fuzzing inputs that cause crashes. The script collects these recorded crash inputs and uses them to create test cases. Finally, the script adds additional test cases using \texttt{STOKE}'s random test case generation algorithm, thereby creating the final set of test cases that are used in \texttt{STOKE}’s synthesis process.


\paragraph{Disassembling and Reassembling IoT Firmware Binaries.}
 \texttt{STOKE} requires the assembly instructions of security-critical basic blocks. \framework~uses \texttt{Ramblr}~\cite{wang2017ramblr} and \texttt{Ddisasm}~\cite{flores2020datalog} to obtain a re-assemblable assembly, disassembling a binary to obtain assembly instructions that can be readily re-assembled into binaries. The obtained re-assemblable assembly allows us to recover the assembly instructions required by \texttt{STOKE} and modify these assembly instructions to incorporate our basic block-level \semanticequiv{s}.
If \framework~fails to recover the assembly instructions of an IoT firmware binary using these disassembling tools, it errors out and stops its diversification process.

 Once we produce basic block-level \semanticequiv{s} for the security-critical basic blocks, \framework~replaces one \semanticequiv~instead of its original basic block in the recovered IoT firmware assembly. After replacing the original basic blocks of Type R and Type M with basic block-level \semanticequiv{s} in the assembly code of the firmware, \framework~uses an off-the-shelf compiler to convert the modified assembly to firmware-level \semanticequiv~binaries, where each binary contains one or more basic block-level \semanticequiv{s}.

\section{Evaluation}
\label{sec:evaluation}


\paragraph{Research Questions.} We formulate three research questions to test the effectiveness of \framework~ across various metrics. \textbf{(R1)} Do \framework's \semanticequiv{s} provide reliable mitigation against the impact of ROP attacks targeting individual IoT binaries, as evidenced by a reduction in successful exploit attempts and diminished resulting damage? \textbf{(R2)} By what percentage does \framework~decrease the success rate of memory error botnet DDoS attacks in comparison to without software diversity? \textbf{(R3)} What are the performance overhead and resource utilization costs associated with using \framework~on an IoT firmware binary, and is this cost acceptable considering the resource constraints of typical IoT devices?

\paragraph{Experimental Setup.}


To evaluate \framework~and to answer our research questions, we design an experimental setup using \ddosim~\cite{obaidat2023ddosim}. \ddosim~consists of three main components: Devs, Attacker, and TServer~\cite{obaidat2023ddosim}. Devs are a collection of Docker containers running real-world IoT binaries. Attacker is a Docker container loaded with scripts and tools to compromise Devs remotely and issue attack commands. TServer is a node in this simulation serving as the target for the compromised Devs' attacks. These components communicate with each other over an NS-3 simulated network.

\ddosim's authors load Attacker with scripts that deliver the Mirai botnet malware binaries to Devs by remotely exploiting buffer overflow vulnerabilities in \texttt{Connman} and \texttt{Dnsmasq} through ROP payloads. Concurrently, they provide TServer with logging tools to record the botnet DDoS attack's intensity~\cite{obaidat2023ddosim}.

We conduct our experiments on a desktop computer with a 4 GHz Intel Core i7 CPU and 32 GB of memory, running Ubuntu 22.04 LTS. 
Similar to \ddosim, we use vulnerabilities in \texttt{Connman} and \texttt{Dnsmasq} for our experiments~\cite{english2019exploiting}.
In addition to the \texttt{Connman} v1.34 CVE-2017-12865 vulnerability that the \ddosim~ paper uses, we also perform experiments on \texttt{Connman} v1.38 memory error vulnerability (CVE-2021-26675), which also exists in v1.34. Similar to the work done by \ddosim's authors, we construct an ROP payload designed to exploit this newer vulnerability and use it to deliver the Mirai malware to the target device~\cite{obaidat2023ddosim}.


Similar to \cite{obaidat2023ddosim}, we employ the \emph{UDPPlain flood} attack, which directs UDP traffic from all compromised Devs (i.e., bots) towards TServer, and we measure the impact of this attack on TServer. We use 100 Devs and monitor the number of Devs that the ROP payload successfully exploits. Following successful ROP exploits, a malware binary is installed on the compromised Devs, which establishes a connection to the Attacker’s command and control server. We then issue a command to the bots, initiating an attack on TServer for 100 seconds (i.e., the attack duration is 100 seconds). For this simulation run, we measure the \emph{overall received data rate} at TServer, calculated as the total traffic received during the attack period from all bots, averaged over time.

\paragraph{Results: Discovered Type R and Type M Blocks.}Table~\ref{tbl:typer_ch_reg} presents our collected statistics for basic blocks containing \texttt{Change Register} and \texttt{Change Memory} gadget classes, identified by \texttt{Radare} and \texttt{Ropper} for \texttt{Dnsmasq}, \texttt{Connman} v1.34, and \texttt{Connman} v1.38. The binaries  have the following sizes: \texttt{Dnsmasq} at 343,728 bytes, \texttt{Connman} version 1.34 at 1,072,688 bytes, and \texttt{Connman} version 1.38 at 1,146,072 bytes.

\begin{table}
\begin{center}
\caption{Type R (Change Register and Change Memory) Basic Blocks}
\label{tbl:typer_ch_reg}
\begin{tabular*}{\linewidth}{@{\extracolsep{\fill}}llllllllll}
\toprule
& \multicolumn{3}{c}{\textbf{Dnsmasq}} & \multicolumn{3}{c}{\textbf{Connman-1.34}} & \multicolumn{3}{c}{\textbf{Connman-1.38}} \\
\cmidrule{2-4} \cmidrule{5-7} \cmidrule{8-10}
& Radare & Ropper & Overlap& Radare & Ropper & Overlap& Radare & Ropper & Overlap\\
\midrule
\multicolumn{10}{c}{\textbf{Change Register Basic Blocks}} \\
\midrule
rdi & 142 & 3 & 3 & 232 & 4 & 3 & 234 & 5 & 2 \\
rsi & 173 & 5 & 4 & 370 & 3 & 2 & 390 & 6 & 4 \\
rdx & 1 & 1 & 1 & 17 & 16 & 11 & 18 & 18 & 14 \\
rcx & 0 & 0 & 0 & 12 & 8 & 6 & 13 & 10 & 6 \\
r8 & 0 & 0 & 0 & 8 & 5 & 3 & 10 & 7 & 5 \\
r9 & 0 & 0 & 0 & 1 & 1 & 1 & 0 & 0 & 0 \\
\midrule
\multicolumn{10}{c}{\textbf{Change Memory Basic Blocks}} \\
\midrule
- & 69 & 54 & 50 & 348 & 212 & 212 & 377 & 231 & 225 \\
\bottomrule
\end{tabular*}
\end{center}
\end{table}

\begin{table}[t]
\begin{center}
\caption{\label{tbl:typer_call_typem} Type R (Call) \& Type M Basic Blocks}
\begin{tabular}{p{1.3cm}p{2cm}p{2cm}p{2cm}}
  \toprule
  & \textbf{Dnsmasq} & \textbf{Connman-1.34} & \textbf{Connman-1.38} \\
  \midrule
  \multicolumn{4}{c}{\textbf{Type R (Call) Basic Blocks}}\\
  \midrule
  Unique & 139   &  469   & 482 \\
  Total  & 1247  &  7579  & 8041 \\
 \midrule
 \multicolumn{4}{c}{\textbf{Type M Basic Blocks}}\\
 \midrule
  - & 494      &  431   & 466 \\
 \bottomrule
\end{tabular}
\end{center}
\end{table}


Table~\ref{tbl:typer_call_typem} presents our collected statistics for basic blocks containing Type R \texttt{Call} gadget class and Type M basic blocks, as identified using \texttt{angr}'s CFG for \texttt{Dnsmasq}, \texttt{Connman} v1.34, and \texttt{Connman} v1.38. In this table, we calculate both the \emph{Unique} (each different identified gadget) and \emph{Total} (every instance a gadget is identified) number of basic blocks for the \texttt{Call} gadget class. 

\paragraph{Results: Type R Diversification.}In our experiments using \ddosim, we generate firmware-level \semanticequiv{s} from \texttt{Dnsmasq} and both versions of \texttt{Connman}, then load these \semanticequiv{s} into Devs. In these firmware-level \semanticequiv{s}, we randomly change between 1-50 Type R basic block-level \semanticequiv{s} in our three binaries.

Our experiments indicate that standard ROP exploits occasionally succeed in exploiting our firmware-level \semanticequiv{s} due to the abundant availability of ROP gadgets. To illustrate this finding using \texttt{Dnsmasq}, an attacker can exploit up to 142 gadgets to assign a value to the \texttt{\%rdi} register. When Devs are exploited due to these reasons, they become bots that can be controlled by the command and control server to participate in the DDoS attack. Our obtained data from these experiments reveals a surge in the average received data rate to 18161348.604 bps when all Devs turn into bots (i.e., all Devs are exploited), as shown in Figure~\ref{fig:throughput_morph_R}.

Figure~\ref{fig:throughput_morph_R} also shows that the average received data rate is higher than setups with fewer active bots (less than 100 Devs are compromised) in some of our experiments. Experiments with fewer bots typically involve basic block-level \semanticequiv{s} that replace key registers, such as \texttt{\%rdx}, \texttt{\%rcx}, and \texttt{\%r8}. The reduced number of bots in these experiments is primarily due to two factors: (1) the ROP exploit in our experiments, responsible for infecting Devs with the Mirai malware, is deeply dependent on these registers---it utilizes the first five argument registers for malware delivery; (2) the number of available gadgets for these particular registers is limited, as highlighted in Table~\ref{tbl:typer_ch_reg}.

Our experiments suggest that to thwart ROP exploits using only Type R basic blocks, we need to diversify all basic blocks within a specific Type R gadget class (such as \texttt{Change Register}). This outcome stems from an attacker's need to write values from the ROP payload (e.g., from the stack in stack-based buffer overflows) to a register before invoking functions that serve their malicious intent---since registers are pivotal for passing arguments to functions. For e.g., an attacker might aim to pass the address of the string "\texttt{/bin/sh}" into the \texttt{\%rdi} register to invoke a shell using the \texttt{system} \texttt{libc} function. By preventing the attacker from employing any of these \texttt{Change Register} gadgets, we can sufficiently counteract the ROP exploit. This rationale is similarly applicable to other gadget classes.

As detailed in our adversarial model in Section~\ref{sec:intro}, we do not assume that Devs employ ASLR. Given this context, an attacker can easily leverage \texttt{Change Register} gadgets to execute a simple \texttt{return-to-libc} exploit, where the attacker utilizes \texttt{libc} functions and manipulates them by passing various arguments. In light of this, the optimal countermeasure is to diversify all Type R \texttt{Change Register} basic blocks, which effectively neutralizes both \texttt{return-to-libc} and ROP exploits.

\pgfplotsset{ignore legend/.style={every axis legend/.code={\let\addlegendentry\relax}}}
\begin{figure}[ht]
\begin{minipage}{\linewidth}
\begin{center}
\begin{tikzpicture}
\begin{axis}[
    xlabel=Number of Devs,
    ylabel=Average received data rate (bps),
    width=7.8cm,height=5cm,
    legend pos=north west,
    legend style={nodes={scale=0.7, transform shape}, anchor=north west,draw=black,fill=white,align=left},
    try min ticks = 5,
    label style={font=\scriptsize},
    ]
\addplot[color=blue!80, dotted, mark=o, every mark/.append style={solid}] coordinates {
(100, 18161348.604)
(75, 14107631.045)
(50, 11972449.184)
(25, 7369984.317)
(0, 0)
};

\end{axis}
\end{tikzpicture}
\caption{Results of diversifying Type R \semanticequiv{s}}
\label{fig:throughput_morph_R}
\end{center}
\end{minipage}
\end{figure}
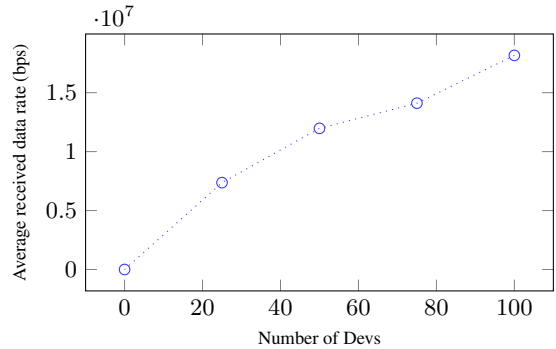

\paragraph{Results: Type M Diversification.} Similar to our Type R experiments using \ddosim, we generate firmware-level \semanticequiv{s} for \texttt{Dnsmasq} and both versions of \texttt{Connman}, then load these \semanticequiv{s} into Devs. In these firmware-level \semanticequiv{s}, we randomly change between 1-50 Type M basic block-level \semanticequiv{s} in our three binaries. 

Initial observations show that diversification of Type M basic blocks in our binaries is not effective in countering ROP exploits, even when substituting known vulnerable Type M basic blocks in these binaries with alternate \semanticequiv{s}.
One potential explanation for this arises from the structure and referencing pattern of these binaries. Specifically, the binaries under examination define the destination buffer (which is overflowed during the attack) in a function, say \texttt{defBuffer()}, distinct from the vulnerable function, \texttt{overflowFunction()}. Due to this separation, \texttt{overflowFunction()} accesses the buffer using an absolute stack address. This behavior contrasts with the use of dynamic address computations relative to the \texttt{\%rsp} register, as seen in instructions like \texttt{movq \%rsi,(\%rsp)}. Given this specific scenario, any modifications or alterations to the stack's structure, as illustrated in our motivating example in \S\ref{sec:selection}, leave the vulnerable buffer unaffected. This situation arises when the buffer is not defined as a local variable within the domain of the function containing Type M basic blocks.

We believe the results from our limited dataset does not indicate ineffectiveness of Type M diversification for security. More experimentation is necessary to offer a definitive conclusion and to explore refining the definition or constraints of Type M.


\paragraph{Discussion.}
Our experimental results support the hypothesis that \framework's \semanticequiv{s} provide reliable mitigation against ROP attacks on individual IoT binaries \textbf{(R1)}. Evidence of this is seen in the reduction in the number of successful exploit attempts, from 100 when no diversification is applied to 0 bots when all Type R change register basic blocks are diversified. 
Towards \textbf{(R2)}, the observed decrease in the average received data rate at TServer when \framework~is applied indicates that \framework~ decreases the success rate of memory error botnet DDoS attacks compared to without using \framework. 
Towards \textbf{(R3)}, as mentioned earlier, we guide \texttt{STOKE} to generate \semanticequiv{s} that are equal to or smaller in size than the original basic blocks. This ensures that \framework~does not increase the binary size, thereby eliminating overheads related to storage capacity. Although we do not directly evaluate performance costs, we infer that diversifying only a limited number of basic blocks allows minimal impact on execution speed.







\paragraph{Limitations.}
First, our evaluation is limited to \texttt{Connman} and \texttt{Dnsmasq}. While these binaries are widely used in IoT devices, and therefore provide a representative sample, we are currently extending our experimental results to a wider set. Second, \framework~currently only supports x86 binaries, a result of employing \texttt{STOKE} as its core diversification engine. However, to cater to a broader range of architectures, we are working on using QEMU and integrating it into \framework. QEMU can facilitate the translation of selected basic blocks across architectures, thereby increasing \framework's versatility. Finally, our localized data flow analysis may lack a comprehensive analysis of global program properties.

Our evaluation of \framework~is confined to three binaries (two versions of \texttt{Connman} and one version of \texttt{Dnsmasq}) due to the necessity of using working exploits within \texttt{DDoSim}'s Docker-based simulation environment. Developing such exploits demands significant manual effort and an in-depth understanding of the target program~\cite{english2019exploiting,suciu2022expected}. For example, a comprehensive analysis by RAND Corp. revealed that creating a fully functional exploit may take up to 955 days, highlighting the extensive time and effort required for this process~\cite{ablon2017zero}. We are currently in the process of extending our experiments to other binaries and vulnerabilities.

\section{Related Work}
\label{sec:related}

Several works in the literature apply \emph{software diversity} and \emph{randomization techniques} to traditional systems, aiming to enhance their security posture~\cite{wartell2012binary,hiser2012ilr,larsen2018continuing,jelesnianski2020mardu,wang2020framework,jiang2023jiujitsu,el2023ng,berlakovich2023r2c}. These approaches effectively complicate exploit development and deployment by introducing unpredictability and variability in software behavior (cf.,~\cite{larsen2014sok}). However, applying these techniques to IoT devices presents challenges due to various factors, such as the reliance on specific hardware features (e.g., virtual memory and segmentation), significant overheads, or the need for virtual machines and runtime rewrites, which are often impractical for IoT devices~\cite{cficare}. Conversely, there has been considerably less work directed towards adapting and developing similar techniques specifically for the IoT domain~\cite{shi2022harm,luo2022faslr,shen2022randezvous,tsoupidi2023thwarting,jang2023effective}. These IoT-focused approaches also encounter limitations, such as overheads and deployment constraints, and often require access to source code, hindering their practical application in real-world IoT environments~\cite{wang2020quantitative,luo2022faslr,shi2022harm}.

The work presented by Pappas et al.~\cite{pappas2012smashing} is similar to our work in \framework, in which the authors utilize a randomization approach to strengthen binary applications against ROP attacks. Their method incorporates narrow-scope modifications, such as substituting an instruction with its functionally equivalent counterpart, aiming to eliminate and probabilistically modify as many ROP gadgets in a process's address space as possible. Mainly, these modifications target executable segments that may contain unintended instruction sequences~\cite{gulwani2017programming}. These sequences, due to the x86 instruction set's complexity, can transform into entirely different instructions if they are decoded from a byte other than the first. Despite the similarity with \framework's Type R basic block diversification, their techniques may not fully neutralize traditional gadgets or be suitable for certain binary sections. For instance, reordering instructions may not work if a sequential execution is necessary. \framework, in contrast, uses \texttt{STOKE} to synthesize semantically identical code sections, mitigating both intended and unintended ROP gadgets. In addition, an informed ROP exploit system with its narrow-scope modifications may circumvent these randomization techniques. For instance, an attacker can load identical values into all registers despite reordering of \texttt{pop} instructions.

\section{Conclusion \& Future Work}
\label{sec:conclusion}

We develop \framework, a software diversity approach designed to counteract ROP exploits in Linux-based IoT devices, without the need for source code, and without requiring special hardware or extensions. Built on top of \texttt{STOKE}, a stochastic optimizer, \framework~achieves diversity by creating functionally equivalent but syntactically different rewrites of an input IoT firmware binary. We develop several technical extensions to \texttt{STOKE}, in order to meet the specific needs of targeting security of IoT binaries. Preliminary results promise \framework's effective mitigation against ROP attacks, with minimal binary size and performance overhead.

At present, we are expanding our experimental results to include more binaries and vulnerabilities, which we plan to publish in future work. We also plan to develop a formal methods based  technique for verifying the semantic equivalence of rewrites.

\bibliographystyle{splncs04}
\bibliography{references}

\end{document}